\documentclass[onecolumn,showpacs,superscriptaddress,aps,prl,twocolumn]{revtex4-2}

\usepackage{graphicx}
\usepackage{amsmath}
\usepackage{amsfonts}
\usepackage{dcolumn}
\usepackage{bm}
\usepackage{soul}
\usepackage[normalem]{ulem}
\usepackage{color}
\usepackage[caption=false]{subfig}
\usepackage[T1]{fontenc}
\usepackage{float}
\usepackage[12h=true]{scrtime}
\usepackage[dvipsnames]{xcolor}
\usepackage{cancel}
\newcommand{\B}[1]{{\bm{#1}}}

\newcommand{\C}[1]{{\mathcal{#1}}}
\DeclareMathOperator{\sech}{sech}
\begin{document}
%\title{Putting to Test a Nonlinear Theory of Shear Banding in Amorphous Solids}
%\title{\color{magenta}{Nonlinear Topological Screening as a basis for shear banding in Amorphous Solids}}
\title{Shear Banding in Amorphous Solids as a Nonlinear Screened Soft Mode Instability}
%\title{Shear Bands in Amorphous Solids: Nonlinear Topological Screening}
%%%%%%%%%%%%%%%%%%%%%%%%%%%%%%%%%%%%%%%%%%%%%%%%%%%%%%%%%%%%%%%%%%%%%%%%%%% 
\author{Yang Fu} 
\affiliation{Hangzhou International Innovation Institute, Beihang University, Hangzhou, 311115, China}
\affiliation{Institute of Theoretical Physics, Chinese Academy of Sciences, Beijing 100190, China}
\author{Yuliang Jin} 
\affiliation{Institute of Theoretical Physics, Chinese Academy of Sciences, Beijing 100190, China}
\affiliation{School of Physical Sciences, University of Chinese Academy of Sciences, Beijing 100049, China} 
\author{ Avanish Kumar} 
\affiliation{Sino-Europe Complex Science Center, School of Mathematics,
	North University of China, Shanxi, Taiyuan 030051, China}
\author{Itamar Procaccia}
\affiliation{Sino-Europe Complex Science Center, School of Mathematics,
	North University of China, Shanxi, Taiyuan 030051, China}
\affiliation{Hangzhou International Innovation Institute, Beihang University, Hangzhou, 311115, China}
\affiliation{Department of Chemical Physics, the Weizmann Institute of Science, Rehovot 7610001, Israel.}

\date{\today}

%%%%%%%%%%%%%%%%%%%%%%%%%%%%%%%%%%%%%%%%%%%%%
\begin{abstract}
Shear banding is a well-known and widespread instability in strained solids: under external strain, the deformation localizes along a line in two dimensions or a plane in three dimensions. Developing a proper theoretical description of this phenomenon is key to understanding mechanical failure in solid materials. Very recently, a nonlinear theory extending classical elasticity to include plastic deformations as topological charges was proposed, offering detailed predictions on the nature and consequences of the shear-banding instability. The theory derives a Hessian operator whose lowest eigenvalue vanishes at the onset of instability, and the corresponding critical eigenmode describes the displacement field across the shear band. The resulting soft mode possesses the selected localization scale and
subsequently saturates into a finite-width shear band. The aim of this Letter is to examine this theory numerically, establishing the role of topological screening and nonlinear instability as the mechanisms governing shear banding during athermal quasistatic deformation. We show that the displacement profile around the shear band is directly determined by the screening parameter and the nonlinear coefficient, thereby quantitatively verifying the theoretical predictions. Our results demonstrate that shear banding differs fundamentally from fracture: it arises from a nonlinear instability of an elastic field screened by plastic deformations. This establishes topological screening as the essential mechanism governing shear banding in amorphous solids.
\end{abstract}
%%%%%%%%%%%%%%%%
%%%%%%%%%%%%%%%%
\maketitle

{\bf Introduction:} 
Shear banding is an important instability, in which a solid material responds to external shear strain
by localizing the strain along a thin line in two dimensions, or along a plane in three dimensions \cite{20BLLVP,18OBBRT}. 
Understanding the mechanism of shear banding is key to explaining failure in a host of materials, from metallic glasses \cite{81DS} to complex fluids \cite{16DFML} — a research topic of great importance for the design of new materials and for exploiting existing ones under extreme conditions.
 As a consequence, localization of deformation has been the focus of  intense research activity since the middle of the 20th century \cite{75RR,80RR}. Recently, advances in direct numerical simulations of amorphous solids under shear provided excellent demonstrations of shear banding taking place under quasistatic conditions \cite{jin2017exploring, 18OBBRT}, with control over the amount of ductility in the studied media using the very useful ``swap'' Monte-Carlo algorithm \cite{GrigeraParisi01}. 

One could expect that a problem so fundamental and so widespread would be approached using a nonlinear theory that, once linearized, would provide an operator with an eigenvalue approaching zero at the shear-banding instability, with an eigenfunction that directly describes the profile of strain across the shear band. In fact, such a theory was lacking until very recently. Apparently the reason for the lack of an appropriate theory was that elasticity theory, which is the natural candidate for such an endeavor, was not built to properly take into account plastic events which are fundamental to the mechanical response of amorphous solids \cite{11HKLP}. Thus attempts to understand shear banding relied either on elasticity theory, neglecting the crucial role of plastic deformations, or on ad-hoc models like the so-called ``elasto-plastic models'' which are basically cellular automata depending on ``rules'' rather than on fundamental theory \cite{18NFMB}.

In recent years, the realization that plastic events create topological charges in the displacement field of the constituents of amorphous solids provided the 
necessary point of view for the development of a theory \cite{21LMMPRS,22MMPRSZ}. Focusing on two dimensions for simplicity, a fundamental plastic event appears as a quadrupolar displacement field, being the `cheapest' way to release stress locally. 
This topological charge is also known
 as an ``Eshelby inclusion''~\cite{eshelby1957determination}.
 %, this topological charge becomes a quadrupolar field when an avalanche of plastic events takes place. \YJ{Is this correct? The field is quadrupolar if the charge is isolated - if there is an avalanche of plastic events, the quadrupolar fields from single events will be mixed, distorted and becoming irregular.}
When the quadrupolar field is not uniform, its gradients define a dipolar field \cite{22BMP}, and it was shown that this field brings in screening and screening lengths that do not exist in classical elasticity theory \cite{25KPS}. 
Substantial effort has been devoted to understanding the consequences of these topological fields~\cite{jin2024intermediate, fu2025long, fu2025analytic, livne2025continuum}. This has led to the emergence of ``anomalous elasticity'' as the key ingredient in a theory that describes and predicts the mechanical response of amorphous solids to external strain, with displacement fields that differ fundamentally from those expected from classical elasticity alone.  (The reader should note that elasto-plastic models fail to generate these crucial toplogical charges \cite{26TOB} and therefore are unable to reproduce our results.) 
Very recently the theoretical efforts culminated with a nonlinear equation for the displacement field of a material under external stress \cite{26KP}. This nonlinear theory provides the long-wished-for linearized Hessian operator that exhibits a vanishing eigenvalue at the shear banding instability, and an eigenfunction that is supposed to describe quantitatively the properties of the shear band, including its width and amplitude, depending on how ductile or brittle the system is. 
In this Letter, we confront the above-mentioned  nonlinear theory with numerical simulations, revealing that
topological screening lies at the heart of the instability that creates shear bands.

{\bf Predictions from the nonlinear theory:} Full details of the nonlinear theory, including its derivation and predictions, can be found in \cite{26KP}. The reader is advised to consult that paper as background for the present Letter. Here we briefly sketch the main aspects. 

The fundamental field to study is the displacement field, from which one can derive the strain and the stress, using linear or nonlinear relations. 
The derivation culminates in an equation for the displacement field $\B d(\B r)$ where $\B r$ is a point in 2-dimensional space. 
The displacement field determines the strain field according to the nonlinear geometric relation \cite{LandauElasticity}
\begin{equation}
	u_{\alpha\beta}= \frac{1}{2}\left[\partial_\alpha  d_\beta + \partial_\beta d_\alpha  \right] + \frac{1}{2} \partial_\beta d_\gamma\partial_\alpha d_\gamma \ .
	\label{defunl}
\end{equation} 

Knowing the strain field associated with any plastic event, including an avalanche of events, we can extract the quadrupolar field. 
We first compute the nonaffine strain $\mathbf{u}^{\mathrm{NA}}$ by subtracting the affine part, and decompose it as a unit tensor and a quadrupolar traceless symmetric tensor 
$
	\mathbf{u}^{\mathrm{NA}} = m \mathbf{I} + \mathbf{Q},
$
where $\mathbf{I}$ is the identity tensor, and $\mathbf{Q}$ is the quadrupolar field. 
Once the quadrupolar tensor field is at hand, the dipolar vector field is obtained as its gradient, 
\begin{equation}
	P^\alpha = \partial_\beta Q^{\alpha\beta}.
	%\YF{P_\alpha = \partial_\beta Q_{\alpha\beta}.}
    \label{Eq:Constitutive}
\end{equation}

Importantly, a linear theory shows that the existence of a dipolar field results in screening, with generally two screening parameters, $\kappa_e$ and $\kappa_o$, which change the solutions for the displacement field from classical elasticity predictions \cite{25CSWDM,24FHKKP}.
They also appear, to leading order, in a linear constitutive relation between the dipolar field and the displacement field,  
\begin{equation}
	P_\alpha = -\Gamma_{\alpha\beta} d_\beta,
	\quad
	\Gamma_{\alpha\beta} = \kappa_e^2 \delta_{\alpha\beta} + \kappa^2_o \epsilon_{\alpha\beta},
	\label{Pind}
\end{equation}
where $\epsilon_{\alpha\beta}$ is the Levi-Civita symbol. 
We find that, in the present case of shear banding, $\kappa_o$ does not play an important role, whereas $\kappa_e$ is crucial, without which there is no shear banding. Thus, we set $\kappa_o = 0$ below. 

The nonlinear theory of Ref.~\cite{26KP} considers a 2-dimensional amorphous material that has been strained quasistatically until a background stress $\B \Sigma$ builds up.  At a certain critical value of this background stress, a plastic event takes place, accompanied by a stress drop $\B \sigma$.  The Lagrangian takes the form
\begin{eqnarray}
	\mathcal{L} &=& \frac{1}{2}(\Sigma_{\alpha\beta}u_{\alpha\beta} + \sigma_{\alpha\beta}u_{\alpha\beta}) + \Gamma_{\beta}^{\alpha}d_{\alpha}P^{\beta} + \frac{1}{2}\Lambda_{\alpha\beta}P^{\alpha}P^{\beta}\nonumber\\ &+&  \frac{1}{4} G_{\alpha \beta \gamma \delta} P^{\alpha}P^{\beta}P^{\gamma} P^{\delta}.
    \label{eq:Lagrangian}
\end{eqnarray}
Minimization of this Lagrangian yields a nonlinear  constitutive relation,
{
\begin{equation}
	P_\alpha = -\Gamma_{\alpha\beta} d_\beta+G_{\alpha\beta\delta\gamma}d_\beta d_\delta d_\gamma,
\label{eq:nonlinear}
\end{equation}}
and a nonlinear equation for the displacement field,
\begin{equation}\label{finalNL}
\Sigma_{\alpha\beta}(d_{k,\beta})_{, \alpha} + \sigma_{\alpha k, \alpha}   +   ({\sigma}_{\alpha \beta}d_{k,\beta})_{, \alpha} = -L^{k\zeta} d_{\zeta} + T^{kpqr}d_p d_q d_r \ ,	
\end{equation}
where the notation $d_{k,\beta}$ means $\partial_\beta d_k$, and the tensor $G_{\alpha\beta\delta\gamma}$, $L^{k\zeta}$ and $T^{kpqr}$ are determined in the theory.
%\YJ{We should unify the notations in the paper: $d_{k,\beta}$ means $\partial_\beta d_k$. For example, see Eq. (1) }

The instability resulting in the formation of a shear band is identified with the existence of a soft mode of an appropriately defined Hessian, allowing us to project the displacement field onto that solution, thereby reducing the problem to a scalar function of a single variable.  
Accordingly, one rewrites Eq.~(\ref{finalNL}) for a scalar function of one variable.  Next one can construct an energy functional whose Euler-Lagrange equations are the same as the aforementioned equation, thereby enabling the explicit determination of the Hessian whose soft eigenfunction identifies the shear-band solution. 
In other words, we are seeking a solution for the displacement field in the form,
	$d_k(\mathbf{x}) = e_k \, f(\xi)$.
The unit vector $\mathbf{e}$ is in the direction of displacement of material points, while the scalar function $f(\xi)$ specifies the magnitude of their displacement as a function of the $\xi$-coordinate perpendicular to the band.
Note that this function can be positive or negative. 
The equation for $f(\xi)$ thus takes the form: 
\begin{eqnarray}
	\big(\mu+\Sigma\sin(2\theta)\big) f''(\xi) &+& \frac{3}{2}(\lambda+2\mu) f'(\xi)^2f''(\xi) =\nonumber\\& -& \C A f(\xi)+\C B f^3(\xi)\ . \label{beauty}
\end{eqnarray}
Here $\theta$ is the angle of the shear band with respect to the shear principal axis, and $\mu$ and $\lambda$ the standard Lam\'e coefficients.
The parameter $\C A$ is related to the screening parameter according to the theory, 
\begin{equation}
\C A = \mu \kappa^2_e,
\label{Akap}
\end{equation}
and $\C B$ is a nonlinear coefficient.
%, and the parameters $\C A$ and $\C B$ are computed explicitly.

Analyses of Eq.~(\ref{beauty}) provide the following predictions: 

(i) {\it Onset of instability.}
The condition that the Hessian possesses a vanishing eigenvalue imposes an explicit requirement on the parameters for shear banding to occur.
For instance, for a system of width $L$ under %periodic 
zero-value
boundary conditions, this happens when 
\begin{equation}
	\Big(\mu+\Sigma\sin(2\theta)\Big)\left(\frac{\pi}{L}\right)^2=\C A.
	\label{criterion}
\end{equation}
This condition is tested explicitly by simulations in the End Matter.
%Appendix C: we find  $\Big(\mu+\Sigma\Big)\left(\frac{\pi}{L}\right)^2 \propto \mu \kappa^2_e$, consistent with the theoretical prediction apart from a simple prefactor. 

(ii) {\it Orientation of the shear band.}
The effect of the background stress is maximal when $\sin(2\theta)=1$,  giving $\theta = 45^\circ$. Thus the well-known appearance of the instability at $ 45^\circ $ with respect to the principal shear axis is selected~\cite{shang2024yielding}. %by the physics, not imposed by any geometry or coordinates.
Since the instability at $ 45^\circ $ is well established, we will not further examine it here. Instead, we fixed the top and bottom boundaries in our simulation system, in order to facilitate measurement of the displacement profile $f(\xi)$. Consequently, the selected shear band is always oriented along the $x$ direction, and $\sin(2\theta)$ is replaced by one in expressions below.

(iii) {\it Shape of the shear band.}
The critical mode which describes the displacement field around the shear band was found to be
\begin{equation}
\label{FI19}
	f(\xi) = f_0\,\tanh\!\left(\frac{\xi - \xi_0}{\ell}\right) \ ,
	\end{equation}
where $\xi_0$ is the center of the shear band.    
Here the width $\ell$ and magnitude $f_0$ are predicted to be 
	\begin{equation}\label{FI18}
		\ell = \sqrt{\frac{2(\mu + \Sigma)}{\C A}} ,
	\end{equation}
	\begin{equation}
		f_0 = c\sqrt{\frac{\C A}{\C B}} \ ,
		\label{f0}
	\end{equation}
%\YJ{we should introduce $\mu$ somewhere; Eq. (8) is the first time?}
where $c$ is coefficient of the order of unity that depends on the boundary conditions. For periodic boundary conditions $c=1$.
The above predictions are important because they link the onset and properties of a shear band to parameters such as $\C A$ and $\C B$, which appear only in the plasticity screening theory and not in elasticity theory. This conveys an essential message: shear banding cannot occur in the absence of screening. The main purpose of this Letter is to test the key predictions Eqs.~(\ref{FI18}) and~(\ref{f0}) via numerical simulations. 
To summarize, \(\ell\) and \(f_0\) are obtained by fitting the displacement profile \(f(\xi)\) to Eq.~(\ref{FI19}); \(\mu\) and \(\Sigma\) are derived from the stress-strain curves; and \(\mathcal{A}\) and \(\mathcal{B}\) are extracted from the constitutive relations. Then Eqs.~(\ref{FI18}) and~(\ref{f0}) are explicitly examined using these numerical values.

%%%%%%%%%%%%%%%%%%%%%%%%%%%%%%%%%%%%%%%%%%%%%%%%%%%
\begin{figure}[h!]
\includegraphics[width=0.85\linewidth]{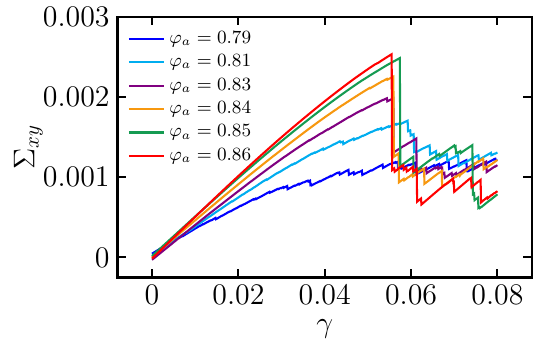}
\caption{The background shear stress $\Sigma_{xy}$ versus strain $\gamma$ curves with different initial annealing conditions ($\varphi_a$), for $N=16000$ Hertzian particles.
%\YF{I will use WCA data.}
%(Panel (a)) and system sizes (Panel (b)).$N=16000$ in (a) and $\varphi_{a}=0.84$ in (b).
}
\label{stvsst_A}
\end{figure}
%%%%%%%%%%%%%%%%%%%%%%%%%%%%%%%%%%%%%%%%%%%%%%%%%%%%

%%%%%%%%%%%%%%%%%%%%%%%%%%%%%%%%%%
\begin{figure}[h!]
	\includegraphics[width=\linewidth]{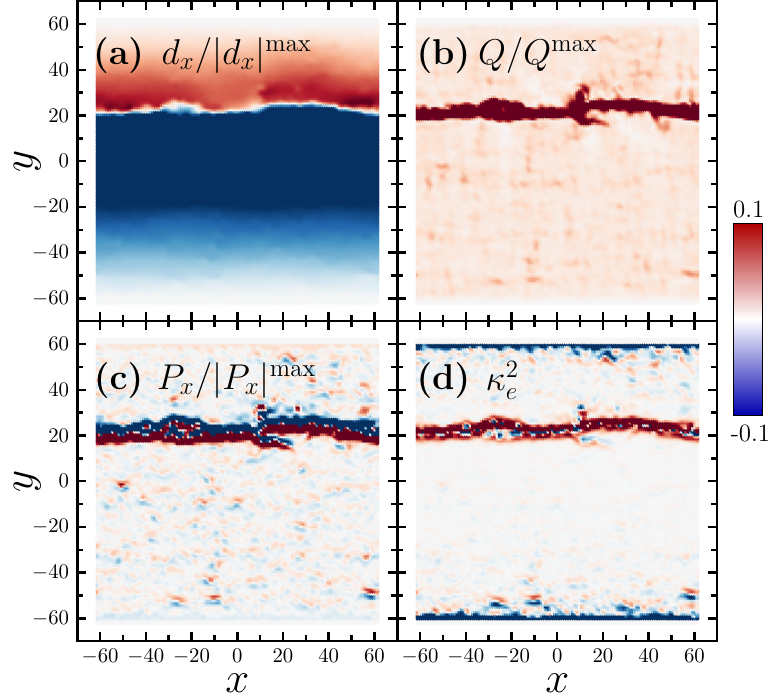}
	\caption{
		Panel (a): the normalized displacement component parallel to the shear band.
		Panel (b): the normalized quadrupole charge, $Q$, defined as $Q=\sqrt{Q_{11}^2+Q_{12}^2}$ {from the components of the quadrupole tensor}.
		Panel (c): the normalized dipole component, $P_{x}$, defined as $P_{x}=\partial_x Q^{xx}+\partial_y Q^{xy}$.
		Panel (d): $\kappa_e^{2}$ calculated based on Eq.~(\ref{kappa_e}).}
	\label{results}
\end{figure}
%%%%%%%%%%%%%%%%%%%%%%%%%%%%%%%%%%%%%%%%%%%%%%%%%%%%%%

{\bf Simulation results:}
We perform athermal quasistatic shear simulations of polydisperse, purely repulsive frictionless spheres in two dimensions. We study three types of particle interactions: harmonic, Hertzian, and Weeks--Chandler--Andersen (WCA) potentials (see Appendix A).
Fig.~\ref{stvsst_A} shows a typical plot of the accumulated stress $\Sigma_{xy}$ versus the strain $\gamma$. 
The data illustrate the dependence on annealing, characterized by the indicated area fraction $\varphi_a$ of the mother liquid prior to compression into the solid phase.
%, which denotes the density of the mother liquid from which compression into the solid phase is initiated. 
As $\varphi_a$ increases (i.e., with a higher degree of annealing), yielding becomes more brittle, as indicated by a larger stress drop~\cite{18OBBRT}. 
Below, we focus on the largest stress-drop event in each sample (see Appendix B for a discussion of other plastic events).
The particle displacement field $\B {d}=d_x \hat{x} + d_y \hat{y}$ provides a direct description of the shear band.
The displacement field is mapped onto a grid using a Gaussian kernel with the grid spacing and Gaussian variance equal to one particle diameter. The plot of $d_x(x,y)$ (panel (a) of Fig.~\ref{results}) shows a sharp change in $d_x(y)$ between the upper and lower half-planes, characteristic of a shear-band event. 

As explained above, having the displacement field we can extract the strain field and from it the quadrupolar field, which is shown in panel (b) of Fig.~\ref{results}. 
The gradient of the quadrupolar field, which is the dipolar field, is shown in panel (c). 
As expected, all these important fields are concentrated along the shear band. 
Finally, using the linear constitutive relation Eq.~(\ref{Pind}), we can extract a pointwise value of the screening parameter $\kappa_e$ according to (see panel (d) of Fig.~\ref{results}) 
\begin{equation}
	\kappa_e^2 = -\frac{d_x P_x + d_y P_y}{d_x^2 + d_y^2}.
	\label{kappa_e}
\end{equation}
The results in Fig.~\ref{results} visualize  the direct correspondence between the shear band and plastic screening. 
%again the conclusion below that without screening there is no shear band.
\begin{figure}
    \includegraphics[width=0.75\linewidth]{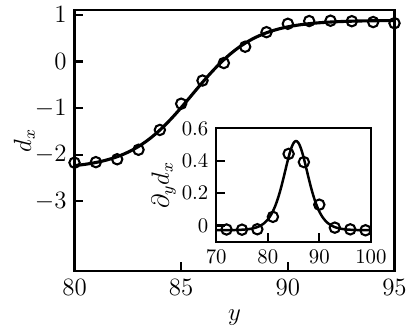}
	\caption{
    Fitting the simulation profile of $d_x(y)$ to the theoretical prediction Eq.~(\ref{FI19}) (line). 
    Inset: comparing the data of shear localization to the theoretical prediction Eq.~(\ref{d_x,y}) with a trivial shift constant.
    } 
		\label{fits}
\end{figure}
%%%%%%%%%%%%%%%%%%%%%%%%%
%%%%%%%%%%%%%%%%%%%%%%%%%%%%%%%%%%%%%%%%%%%%%%%%%%%
\begin{figure}
    \includegraphics[width=1\linewidth]{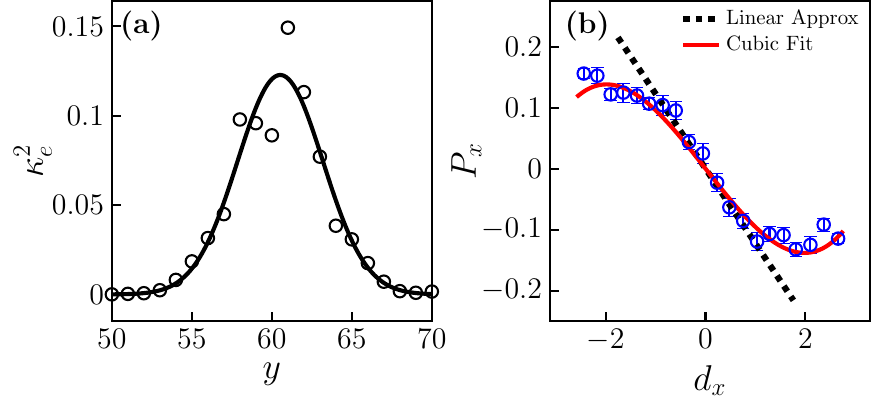}
	\caption{
    Panel (a): The profile of $\kappa_e^2$ in the shear band is fitted by a Gaussian function to determine the peak value $\kappa_{e, {\rm max}}$ {, with $\kappa_e^2$ calculated using the linear constitutive relation Eq.~(\ref{kappa_e})}.
    Panel (b): $P_x$ vs. $d_x$ in the shear band {within $58<y<63$}, fitted to Eq.~(\ref{eq:nonlinear2}){, giving $\C A/\mu\approx1.04$ and $\C B\approx0.0088 $ (red solid line). The black dashed line represents $\C A/\mu\approx1.2$ obtained from panel (a).} 
    }
		\label{fig:A_B}
\end{figure}
%%%%%%%%%%	
Since $|d_x| \gg |d_y|$ and the variation of displacement parallel to the shear band can be ignored, the horizontal displacement $d_x(y)$ is of primary interest. The $x$-averaged data for $d_x(y)$ is shown in Fig.~\ref{fits}, together with a fit to the theoretical prediction Eq.~(\ref{FI19}). The agreement appears satisfactory. An even better comparison that is free of the effect of boundary conditions is provided by the shear localization
\begin{equation}
\partial_y d_x = f_0 \,\mathrm{sech}^2\!\left(\frac{y-y_0}{\ell}\right)+ {\rm const},
\label{d_x,y}
\end{equation}
with a trivial small shift constant (see the inset of Fig.~\ref{fits}). 
Using  such profiles, we extract both $f_0$ and $\ell$ via fitting.

\begin{figure}[h!]
	\includegraphics[width=0.8\linewidth]{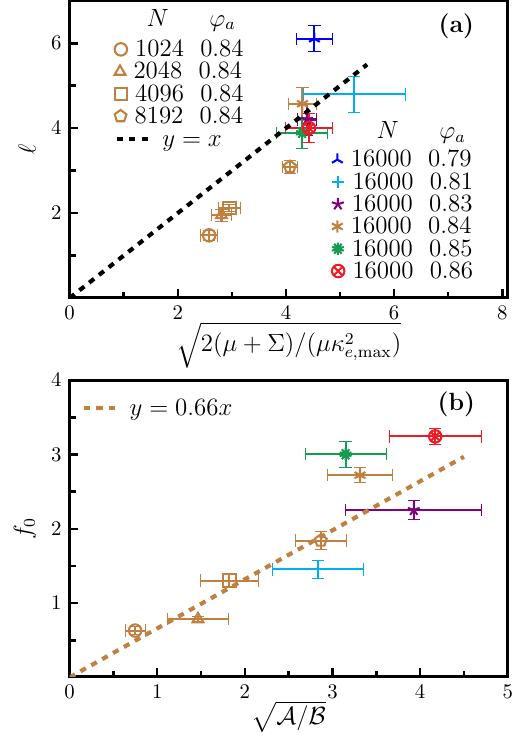}
	\caption{Examination of Eq.~(\ref{FI18}) (panel a), and Eq.~(\ref{f0}) (panel b)  
    using simulation data for WCA systems.
   % \YJ{$ y = 0.6 x$ in (b), we don't need so many digits, the data are not that accurate; change $\kappa_e$ to $\kappa_{e, {\rm max}}$}
    }
	\label{compare}
\end{figure}

Next, we determine $\mathcal{A}$ and $\mathcal{B}$. Figure~\ref{fig:A_B}(a) shows the $x$-averaged $\kappa_e^2$ data. Since we are only interested in the shear band region, we set $\mathcal{A} = \mu \kappa_{e,\text{max}}^2$ using Eq.~(\ref{Akap}), where $\kappa_{e,\text{max}}$ is the maximum $\kappa_e$ at the center of the shear band. To determine $\C B$, the nonlinear constitutive relation Eq.~(\ref{eq:nonlinear}) is reduced to a 1-dimensional relation as an approximation, 
\begin{equation}
P_x = - \frac{\C A}{\mu} d_x + \C B d_x^3.
\label{eq:nonlinear2}
\end{equation}
Fitting simulation data to Eq.~(\ref{eq:nonlinear2}) gives $\C A$ that is consistent with $\mu \kappa_{e,\text{max}}^2$ estimated above, and $\C B$ (see Fig.~\ref{fig:A_B}b).

Using the parameter values obtained from simulations, we examine the theoretical relations given by Eqs.~(\ref{FI18}) and~(\ref{f0}), as shown in Fig.~\ref{fig:A_B} for WCA systems and in Appendix C for harmonic and Hertzian systems. 
We find that Eq.~(\ref{FI18}) works reasonably well, while Eq.~(\ref{f0}) is in excellent agreement with our data, with $c\approx 0.66$ which is consistent with our simulation boundary conditions. In addition, our simulation data in Appendix D explicitly verify the condition given by Eq.~(\ref{criterion}) for the onset of shear-banding instability.

The above numerical tests confirm the key results of the nonlinear theory. The width of the shear band is controlled by the screening length (i.e., inverse screening parameter), $\ell \sim 1/\kappa_e$. In Appendix E we offer an analytic derivation of this important scaling relation for the present case.
In the limit $\kappa_e \to 0$, screening disappears together with the shear band ($\ell \to \infty$). In the other limit, $\kappa_e \to \infty$, the shear band becomes a singular line ($\ell \to 0$) formed by dipolar charges. On the other hand, the amplitude $f_0$ of the shear band depends on both the screening parameter $\kappa_e$ and the nonlinear coefficient $\C B$. In appendix F we show that EP models fail to predict the correct relations between $f_0$ and $\ell$.  Plastic screening and nonlinearity are the origin of shear banding.

{\bf In summary}, we have demonstrated that shear banding in amorphous solids is a screened
soft-mode instability. Dipolar fields screen the elastic
response, resulting in a finite localization scale that is absent from classical
elasticity. Screening modifies the Hessian spectrum, bringing its lowest eigenvalue to zero at yielding, producing a soft mode whose nonlinear saturation forms a finite-width shear band. The key outcome is that the localization scale is not imposed externally but emerges from plastic screening. Our simulations reveal the displacement, quadrupolar, dipolar, and screening fields associated with shear localization and quantitatively confirm the relationships between screening and shear-band structure as predicted by the theory. More importantly, the numerical results demonstrate that screening effects are fundamental to explaining
the phenomenon: shear banding is not fracture, but rather a collective instability of an
elastic field screened by dipolar charges. These results identify plastic screening as the mechanism of shear localization and provide a microscopic continuum description of shear-band formation in amorphous solids.

\bibliography{All_citation}
%\bibliography{ALL,All.Anomalous}

\clearpage

\onecolumngrid
\begin{center}
{\bfseries\Large End Matter}
\end{center}
\twocolumngrid

%{\it{Appendix A: The calculation of $\B P$, $\mathbf{Q}$, $\kappa_e$ \textemdash}} Put Eqs.2-4, Fig.~7 here.

{\it{Appendix A: Simulation methods and models\textemdash}
\label{Appendix_A}}
%%%%%%%%%%%%%%%%%%%%%%%%%%%%%%%%%%%%%%%%%%%%%%%%%%%
The initial configurations are equilibrium states, prepared via the swap Monte Carlo algorithm at an area fraction  $\varphi_a$, below the jamming density $\varphi_{\rm J}$ \cite{GrigeraParisi01}.  The equilibrated configuration is then rapidly compressed  to $\varphi=0.98$, deep in the jammed regime, followed by energy minimization to stabilize the system using the FIRE algorithm~\cite{bitzek2006structural}.
Systems with particle numbers ranging from $N=1024$ to $16000$ are studied.
Particle interactions include harmonic, Hertzian, and Weeks--Chandler--Andersen (WCA) potentials, with increasing stiffness.
%, allowing us to study the effect of stiffness \YF{Discuss its effect in Appendix C? If not, change it to `with increasing stiffness'}.
All particles have unit mass. Thirty  independent samples are generated for each set of parameters.

Simple shear is applied using Lees--Edwards boundary conditions with strain increment $\delta\gamma=10^{-4}$. 
Systems are sheared up to $\gamma_{\rm max}=0.08$ for harmonic and Hertzian systems, and $\gamma_{\rm max} = 0.12$ for the WCA system.
The value of $\gamma_{\rm max}$ is larger than the yield strain $\gamma_{\rm Y}$,
ensuring the occurrence of a large stress-drop event. 
A system-spanning shear band is always observed, as the particles in the top and bottom layers are fixed during energy minimization. 
This explicitly breaks the periodic boundary condition along the $y$ axis and the $\pm45^\circ$ symmetry relative to the principal axes of stress, thus selecting a shear band oriented along the $x$ direction.
Rattlers (particles with $\le 2$ contacts) are removed from mechanically stable configurations.
All simulations are performed using the LAMMPS package~\cite{THOMPSON2022108171}.

\begin{figure}[h!]
\includegraphics[width=1\linewidth]{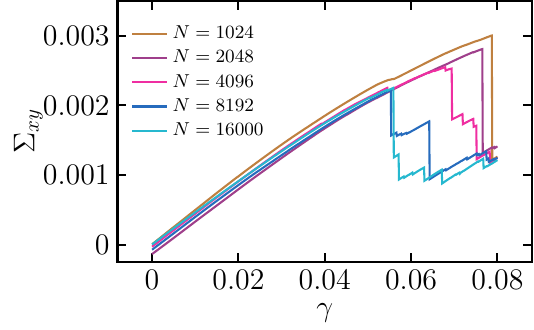}
\caption{
The background shear stress versus strain curves with different system sizes, for Hertzian systems.
%\YF{I will use WCA data.}
}
\label{Fig6}
\end{figure}

Fig.~\ref{Fig6} shows typical curves of accumulated stress $\Sigma_{xy}$ versus strain $\gamma$ for different $N$. 
For each system size, we identify and focus on the largest stress-drop event, which corresponds to the system-spanning shear band.
%For each system size, the largest stress drop can be identified. 
%We mainly focus on shear banding of these large stress-drop events.

%%%%%%%%%%%%%%%%%%%%%%%%%%%%%%%%%%%%%%%%%%%%%%%%%%%%

\begin{figure}[h!]
	\includegraphics[width=1\linewidth]{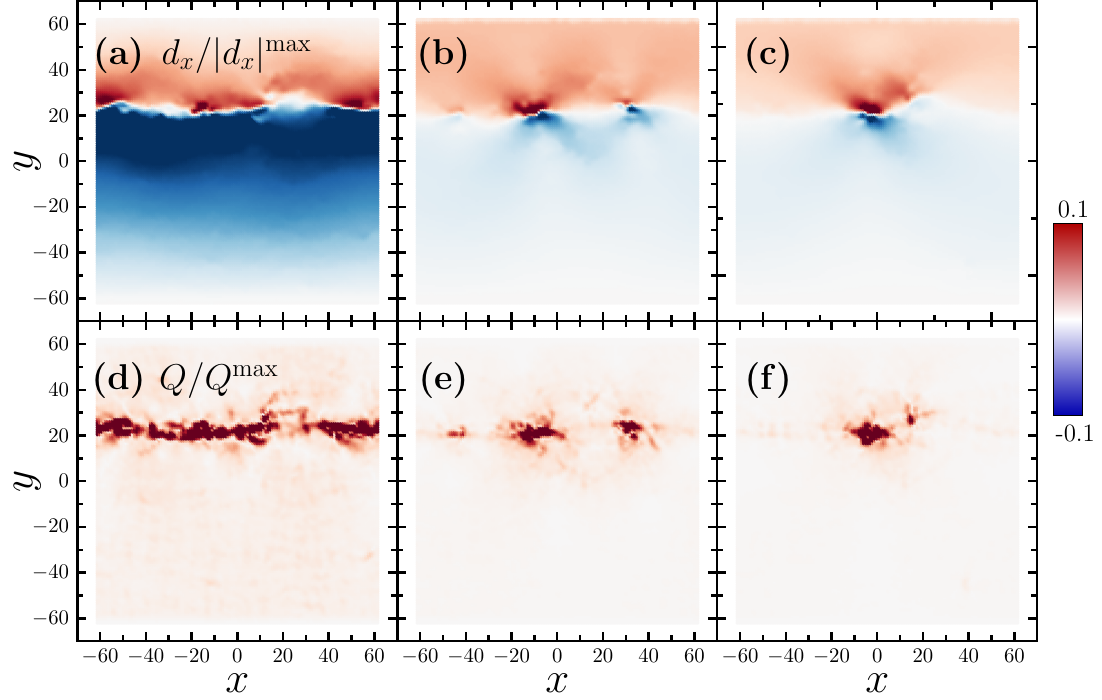}
	\caption{Two typical plastic events (b,c,e,f) after the largest stress drop (a,d).}
	\label{Fig7}
\end{figure}

{\it {Appendix B: Remnants of shear banding in consecutive plastic events after yielding\textemdash}}
%{\it{Appendix D: Instability criterion as a boundary of predicted shear banding width\textemdash}} 
After the largest stress-drop event, plastic events with smaller stress drops continue to occur. Fig.~\ref{Fig7} shows a typical shear band at the largest stress-drop event at the yielding strain $\gamma_{\rm Y}$, along with two typical plastic events occurring after yielding ($\gamma > \gamma_{\rm Y}$). For such events, the theoretical relationships, such as Eqs.~(\ref{FI18}) and~(\ref{f0}),  no longer hold strictly. Interestingly, we find that quadrupolar charges are sparsely distributed along the residual ``scar'' of the original shear band.

%%%%%%%%%%%%%%%%%%%%%%%%%%%%%%%%%%%%%%%%%%%%%%%%%

{\it{Appendix C: Examination of Eqs.~(\ref{FI18}) and~(\ref{f0}) for harmonic and Hertzian systems\textemdash}}
\begin{figure}[h!]
	\includegraphics[width=1\linewidth]{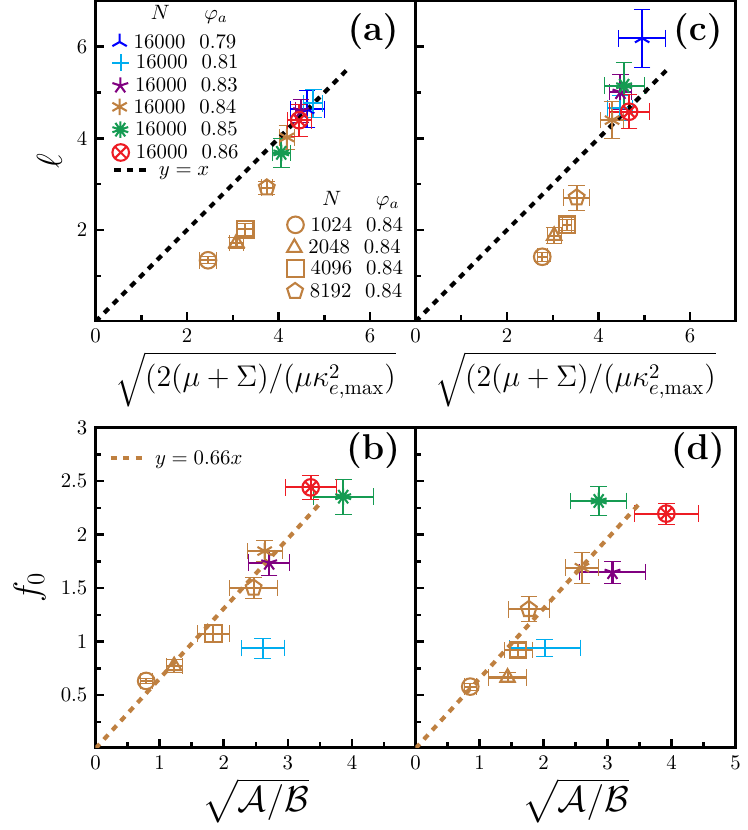}
	\caption{Examination of Eq.~(\ref{FI18}) (panels a,c) and Eq.~(\ref{f0}) (panels b,d) using simulation data for harmonic (left panels) and Hertzian systems (right panels).
    %\YJ{$y=0.66x$?}
    }
	\label{Fig8}
\end{figure}
%%%%%%%%%%%%%%%%%%%%%%%%%%%%%%%%%%%%%%%%%%%%%%%%%
Figure~\ref{Fig8} shows that Eqs.~(\ref{FI18}) and~(\ref{f0}) are consistent with our simulation results for both harmonic and Hertzian systems. Hence, these theoretical results are general and independent of interaction details, such as the range and stiffness of the interparticle potential.\\

\begin{figure}[h!]
	\includegraphics[width=0.8\linewidth]{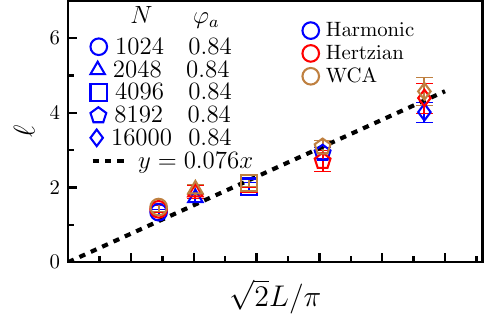}
	\caption{Universal linear relationship  between $\ell$ and $L$ revealed by the simulation data from three simulated systems: harmonic (blue), Hertzian (red), and WCA (yellow) interactions.}
	\label{Fig9}
\end{figure}

{\it {Appendix D: Examination of instability condition\textemdash}} 
To test the instability condition given by Eq.~(\ref{criterion}), one approach is to examine the relationship between $(\mu + \Sigma)(\pi/L)^2$ and $\C A = \mu \kappa_{e, {\rm max}}$. A numerically more direct and robust method is to investigate the relationship
\begin{equation}
\ell = \frac{\sqrt{2}}{\pi} L 
\label{eq:instability}
\end{equation}
derived by combining Eqs.~(\ref{criterion}) and~(\ref{FI18}). Equation~(\ref{eq:instability}) reveals a surprisingly simple relationship between the shear-band width $\ell$ and the linear system size $L$, in which all material-dependent parameters cancel out. Figure~\ref{Fig9} indeed confirms such a simple relation,  $\ell \approx 0.034 L$, for all three simulated systems --- note that $\ell$ and $L$ are not rescaled by any system-dependent parameters in this plot. The numerical coefficient {$0.034$} appears to differ from the theoretical value $\sqrt{2}/\pi$, possibly due to the effects of boundary conditions.
%\YJ{plot all three systems together; Fig. 8c,f and Fig. 9}

%%%%%%%%%%%%%%%%%%%%%%%%%%%%%%%%%%%%%%%%%%%%%%%%%
%%%%%%%%%%%%%%%%%%%%%%%%%%%%%%%%%%%%%%%%%%%%%%%%%

%\YJ{Directly test Eq. (9), don't need to introduce $\ell_{th,2}$}

{\it{Appendix E:  Simplified derivation of $\ell \sim 1/\kappa_e$\textemdash}}
It is useful to gain insight into this inverse relationship through a simplified derivation,  
without going through the detailed calculations in Ref.~\cite{26KP}. We use the fact that the displacement field here is dominated by the critical mode, and has the simple form of Eq.~(\ref{FI19}).  
We can therefore compute explicitly all the fields discussed above, and understand in detail the relationships that are tested here. 
In the 1-dimensional projection approximation, the displacement field $\B d(x,y)$ is described by  $d_x(y) \approx d_x^{\mathrm{NA}} (y)$. 
Then the nonaffine strain tensor basically reads
\begin{equation}
\label{eq:uNA}
	\mathbf{u}^{\mathrm{NA}} = \begin{pmatrix}
		0 & \partial_y d_x/2 \\
		\partial_y d_x/2 & 0
	\end{pmatrix} \equiv \begin{pmatrix}
		0 & u_{xy} \\
		u_{xy} & 0
	\end{pmatrix}	 \ .
\end{equation}

The quadrupolar field is the symmetric traceless tensor that is obtained from the strain field by eliminating the trace.
%Because $\mathbf{u}^{\mathrm{NA}}$ is traceless, we obtain
From $
	\mathbf{u}^{\mathrm{NA}} = m \mathbf{I} + \mathbf{Q}
$, we obtain,
$\mathbf{Q} = \mathbf{u}^{\mathrm{NA}}$,
    %-m \begin{pmatrix}
	%	1 & 0 \\
	%	0 &  1
	%\end{pmatrix} 	
    %=  \mathbf{u}^{\mathrm{NA}},
%\end{equation}
because $\mathbf{u}^{\mathrm{NA}}$ in Eq.~(\ref{eq:uNA}) is traceless. 
%where $m=\rm{Tr}(\B u)=0$ because  $\mathbf{u}^{\mathrm{NA}}$ is traceless.
%, to obtain $\mathbf{Q}= \mathbf{u}^{\mathrm{NA}}$.
Now the dipole field Eq.~(\ref{Eq:Constitutive}) is translated to 
%\begin{equation}
$P_x = \partial_y Q_{xy} = \frac{1}{2} \partial^2_y d_x (y)$.
    %\YF{P_x = \partial_y Q_{xy} = \frac{1}{2} \partial^2_y d_x (y) \ .}
%\end{equation}
We also know the relationship between the dipole field and the displacement field, which in the absence of the odd screening exponent $\kappa_o$ reads
$\B P = -\kappa_e^2 \B d$,
or in the present case
$\frac{1}{2} \partial^2_y d_x = -\kappa_e^2 d_x$.
Performing the differentiation we find
\begin{equation}
	-\frac{f_{0}}{\ell^2} \tanh(y/\ell)\sech^2(y/\ell)=-\kappa_e^2 f_{0}\tanh(y/\ell) \ .
\end{equation}
Canceling the $\tanh(y/\ell)$ on both sides and evaluating at $y=0$, we get
\begin{equation}
	\ell = 1/\kappa_e \ .
    \label{eq:l_kappae}
\end{equation}
The more precise prefactor is in Eq.~(\ref{FI18}), but this simplified derivation highlights the intimate relationship between anomalous elasticity and screening phenomena, and the present analytic theory of shear banding.

\begin{figure}[h!]
	\includegraphics[width=0.75\linewidth]{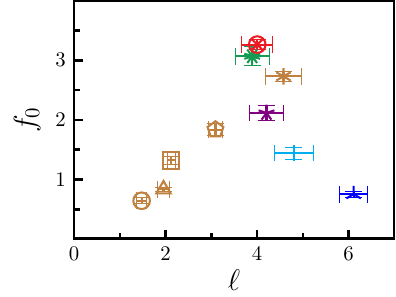}
	\caption{Parametric plot of $f_0$ versus $\ell$ for the same data as in Fig.~\ref{compare}. The symbols are defined in the legend of Fig.~\ref{compare}.}
   % which characterize the strain profile in Eq.~(\ref{d_x,y}), using the same data as in Fig.~\ref{compare} for different system sizes and annealing conditions for WCA systems.}
	\label{Fig10}
\end{figure}

\textit{Appendix F: Comparison with Elasto-Plastic (EP) Scaling--} 
Ref.~\cite{rossi:tel-04241288} (see their Fig.~1.6) reports an empirical finite-size scaling analysis of the shear band profile from the results of an EP model, according to which $f_0 \sim L^{1-\beta}$, $\ell \sim L^\beta$, and $f_0 \sim \ell^{(1-\beta)/\beta}$, with $\beta = 0.61$ in 2D. Using the same data plotted in Fig.~\ref{compare}, we find that our simulation results do not follow the simple relation $f_0 \sim \ell^{(1-\beta)/\beta}$ as observed in the EP model (see Fig.~\ref{Fig10}). Instead, two independent parameters ($\C A$ and $\C B$ in our theory) are required to explain the behavior of $f_0$ and $\ell$ as shown in Fig.~\ref{compare}. The inadequacy of existing models calls for an analytical theory, such as the one developed here. We reiterate that the EP model lacks a predictive analytical description of the shear band.

%Using the same data as in Fig.~\ref{compare}, we plot the relation between the amplitude ($f_0$) and the shear-band width ($\ell$) characterizing the strain profile in Eq.~(\ref{d_x,y})  (similar results for the harmonic and Hertzian systems). As shown in Fig.~\ref{Fig10}, the data does not collapse.This contrasts with the empirical scaling  adopted in the elasto-plastic (EP) model (see the scaling of strain profile in Fig.~1.6 of Ref.~\cite{rossi:tel-04241288}). Instead, as shown in Fig.~\ref{compare}, the present theory quantitatively predicts the shear-band profile by relating $\ell$ and $f_0$ to two independent screening parameters, $\C A$ and $\C B$, providing a predictive capability beyond the empirical EP scaling.

\end{document}